\documentclass[aps,prl,twocolumn,showpacs,superscriptaddress]{revtex4-2}
\usepackage{amsmath,amssymb}
\usepackage{graphicx}
\usepackage{wasysym}
\usepackage{amsfonts}
\usepackage{bm}
\usepackage{enumerate}
\usepackage{color}

\usepackage{url}
\usepackage{hyperref}
\usepackage{epstopdf}
\usepackage{latexsym}
\usepackage[caption=false,singlelinecheck=false]{subfig}
\usepackage{balance}
\usepackage{mathtools}

\DeclarePairedDelimiter\ket{\lvert}{\rangle}
\newcommand{\bea}{\begin{eqnarray}}
\newcommand{\eea}{\end{eqnarray}}
\newcommand{\bs}{\boldsymbol}

\begin{document}
\title{Quantum spin models of commensurate $p$-wave magnets}
	
\author{GiBaik Sim}
\email{gibaik.sim@unimelb.edu.au}
\affiliation{School of Physics, The University of Melbourne, Parkville, VIC 3010, Australia}

\author{Stephan Rachel}
\email{stephan.rachel@unimelb.edu.au}
\affiliation{School of Physics, The University of Melbourne, Parkville, VIC 3010, Australia}
	
\date{\today}
\begin{abstract}
The $p$-wave magnet has emerged as a new type of magnetism exhibiting odd-parity, time-reversal-symmetric spin splitting in momentum space, and has attracted considerable interest as a promising platform for spintronic applications. However, the theoretical understanding of the fundamental mechanism responsible for stabilizing this phase remains limited. In this work, we identify a microscopic interacting model that realizes the $p$-wave magnet as its ground state. We first introduce a Hubbard model and derive the corresponding low-energy spin Hamiltonian. At the classical level, we find that the $p$-wave magnet is stabilized but remains energetically degenerate with competing noncoplanar states. Quantum fluctuations lift this degeneracy, selecting the $p$-wave magnet as the unique ground state. The resulting electronic structure exhibits finite spin accumulation via the Edelstein effect, highlighting the potential of $p$-wave magnetism for spintronic applications. We further discuss the relevance of our theory to quasi-two-dimensional honeycomb magnets such as Ni$_2$Mo$_3$O$_8$. Our findings establish the possibility of spontaneous $p$-wave magnetism.
\end{abstract}

\maketitle

\textbf{\textit{Introduction.---}} The spin texture of electronic bands in momentum space is closely intertwined with the symmetry of magnetic order in real space.\cite{moriya2012spin,vzutic2004spintronics,hayami2018classification,jungwirth2025altermagnetism} In standard ferromagnets, broken time-reversal symmetry lifts spin degeneracy and generates uniform spin splitting across the Brillouin zone, giving rise to an even-parity $s$-wave spin polarization on the Fermi surface.~\cite{katsnelson2008half} Momentum-dependent spin splitting can also emerge in magnets with more complex order. In particular, altermagnets ($d$-wave magnets)~\cite{ahn2019antiferromagnetism,hayami2020bottom,yuan2020giant,vsmejkal2022beyond,vsmejkal2022emerging,leeb2024spontaneous,li2025exploring,PhysRevLett.132.176702,kaushal2025altermagnetism,durrnagel2025altermagnetic,kaushal2026spontaneous} exhibit sign-changing spin polarization in momentum space, reflecting the underlying crystalline symmetry of their collinear magnetic structure.~\cite{vsmejkal2022anomalous,song2025altermagnets,bai2024altermagnetism,liu2025altermagnetism}

Earlier attempts to realize odd-parity, time-reversal-symmetric $p$-wave spin splitting focused on a angular momentum $l$=1 spin Pomeranchuk instability of a Fermi liquid.~\cite{hirsch1990spin,wu2004dynamic,varma2006helicity,wu2007fermi} However, this mechanism was later shown to be prohibited by the spin conservation law.~\cite{kiselev2017limits,wu2018conditions} More recent studies circumvent this constraint by generating $p$-wave spin splitting from noncollinear magnetic states.~\cite{hellenes2023p,chakraborty2025highly,brekke2024minimal} Within this framework, the $p$-wave magnet has been interpreted as a commensurate spiral order whose commensurability restores time-reversal symmetry when combined with lattice translations.

In parallel, significant progress has been achieved on both experimental and theoretical fronts.~\cite{brekke2024minimal,song2025electrical,yamada2025metallic,yu2025odd,chen2024enumeration,xiao2024spin,jiang2024enumeration,10.21468/SciPostPhys.18.3.109,wu2019incommensurate,lu202375,hellenes2023p,chakraborty2025highly,shishidou2018magnetic,stadel2022multiple,yu2025odd,dsouza2025odd,zhu2025floquet,huang2025light,li2025floquet,wnqs-3djt,lin2025odd,leeb2026collinear,li2026p,kim2026odd} Experimentally, $p$-wave magnetic order has been directly confirmed in both insulating and metallic systems: NiI$_2$~\cite{song2025electrical} exhibits an electrically switchable multiferroic $p$-wave state, while Gd$_3$(Ru$_{0.95}$Rh$_{0.05}$)$_4$Al$_{12}$~\cite{yamada2025metallic} hosts a commensurate coplanar spin helix characteristic of a metallic $p$-wave magnet. On the theoretical side, first-principles calculations~\cite{hellenes2023p,chakraborty2025highly,yu2025odd} and symmetry-based classifications~\cite{chen2024enumeration,xiao2024spin,jiang2024enumeration,10.21468/SciPostPhys.18.3.109}, have predicted various forms of odd-parity magnetism in correlated materials. These include the $p$-wave candidate CeNiAsO~\cite{wu2019incommensurate,lu202375,hellenes2023p,chakraborty2025highly} and several Fe-based compounds~\cite{shishidou2018magnetic,stadel2022multiple,yu2025odd,dsouza2025odd} hosting higher-order odd-parity spin textures. Moreover, Floquet-engineered schemes have been proposed to dynamically induce such odd-parity spin splitting.~\cite{zhu2025floquet,huang2025light,li2025floquet,wnqs-3djt}

Despite these advances, microscopic realizations within interacting Hubbard-type or spin models remain largely unexplored~\cite{lee2025inversion,ntsh-ypmc}. Establishing such models is crucial for understanding the fundamental stabilization mechanisms of $p$-wave magnetic order and for being able to access excitation spectra. In this work, we identify Hubbard models that realize the $p$-wave magnet as their ground state in the strong-coupling limit. Starting from a Hubbard model, we derive the low-energy spin Hamiltonian and find that the $p$-wave magnet is stabilized but remains energetically degenerate with competing noncoplanar states at the classical level. To determine the ground state in the quantum regime, we investigate the corresponding spin-1/2 model using matrix product state simulations based on infinite density-matrix renormalization group (iDMRG). We demonstrate that quantum fluctuations lift the classical degeneracy and select the $p$-wave magnet as the unique quantum ground state. Furthermore, we show that the resulting minimal tight-binding model exhibits finite spin accumulation manifested in the Edelstein effect, revealing the potential relevance of $p$-wave magnetism for spintronics applications.

\begin{figure}[]
	\includegraphics[width = 1.0\columnwidth]{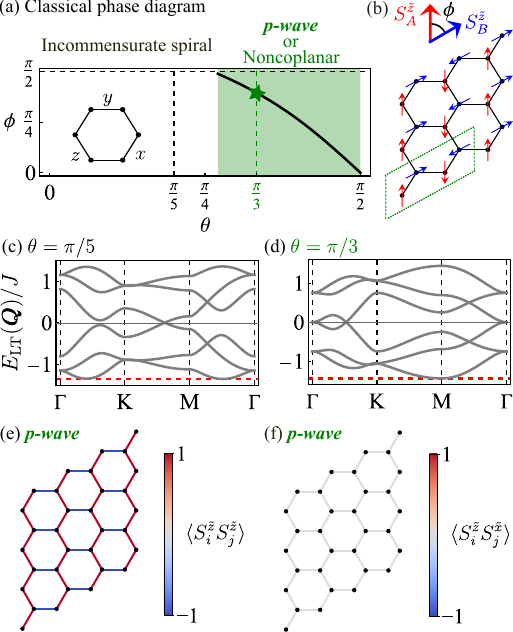}		
	\caption{(a) Classical phase diagram of the model given in Eq.~(\ref{eq:spin}) for $\hat{\bs d}_\gamma = \hat{\gamma}$. For $0.85 \le \theta < \pi/2$, the model stabilizes the $p$-wave magnetic and noncoplanar states, which are energetically degenerate. (b) Spin configuration of the $p$-wave magnet, which is noncollinear. The canting angle $\phi$ between the local spin-polarization axes $S^{\tilde z}_A$ and $S^{\tilde z}_B$ quantifies the noncollinearity between the two sublattices. The magnetic unit cell, indicated by the dashed green outline, is twice as large as the crystallographic unit cell. (c,d) Eigenvalue spectra of the LT interaction matrix for $\theta = \pi/5$ and $\pi/3$. The minimum of the lowest-energy band is located at an incommensurate wave vector for the former and at the M point for the latter. (e,f) Nearest-neighbor spin correlators $\langle S^{\tilde z}_{i} S^{\tilde z}_{j} \rangle$ and $\langle S^{\tilde z}_{i} S^{\tilde x}_{j} \rangle$ for the $p$-wave magnet.}
	\label{fig_1}
\end{figure}

\textbf{\textit{Hubbard model on honeycomb lattice.---}}
We first introduce a variant of the Hubbard model on the two-dimensional honeycomb lattice, which is written as
\small
\bea
\nonumber
H = t \sum_{\langle ij \rangle \in \gamma } 
c_{i}^\dagger 
\big[\cos{\theta} + i \sin{\theta} \, \hat{\bs d}_{\gamma} \! \cdot \! \bs \sigma \big]  
c_{j}
+ U \sum_i n_{i,\uparrow} n_{i,\downarrow},
\\
\label{eq:hubbard}
\eea
\normalsize
where $c_{i}^\dagger \!\equiv\! (c^\dagger_{i,\uparrow}, c^\dagger_{i,\downarrow})$ is the two-component spinor, and the Pauli matrices $\bs \sigma$ act on the spin degree of freedom. The first term describes nearest-neighbor hopping, where $\theta$ quantifies the ratio between spin-independent and spin-dependent hopping amplitudes, and $\hat{\bs d}_{\gamma}$ specifies the form of spin-dependent hopping on the $\gamma = x, y, z$ bonds (see inset of Fig.~\ref{fig_1}(a) for the bond labeling). The second term represents the usual onsite Coulomb repulsion. We consider two representative configurations of $\hat{\bs d}_{\gamma}$:
\bea
\text{(i)}&&\hat{\bs d}_{\gamma} = \hat{\gamma},
\label{eq:d_1}
\\
\text{(ii)}&&\hat{\bs d}_{x} = (1,0,0), \hat{\bs d}_{y} = C_{3z}\hat{\bs d}_{x}, \hat{\bs d}_{z} = C^2_{3z}\hat{\bs d}_{x}
\label{eq:d_2}
\eea
where $C_{3z}$ denotes a threefold rotation by $120^\circ$ about the $z$ axis. In the strong-coupling limit $U/t \!\to\! \infty$, Eq.~(\ref{eq:hubbard}) can be projected onto the subspace with one electron per site, leading to an effective spin Hamiltonian,
\bea
\nonumber
H_s &=& J \sum_{\langle ij \rangle \in \gamma} \big[\cos{2\theta}\, \bs{S}_i \!\cdot\! \bs{S}_j
+ \sin{2\theta}\, \hat{\bs d}_{\gamma} \!\cdot\! (\bs{S}_i \!\times\! \bs{S}_j)
\\
&+& (1 - \cos{2\theta}) (\hat{\bs d}_{\gamma} \!\cdot\! \bs{S}_i)(\hat{\bs d}_{\gamma} \!\cdot\! \bs{S}_j)
\big],
\label{eq:spin}
\eea
where $J=4t^2/U$.~\cite{macdonald1988t,shekhtman1992moriya} The first term denotes the isotropic Heisenberg exchange, while the second and third terms represent the Dzyaloshinskii–Moriya (DM) and Kitaev-type anisotropic interactions, respectively. Hereafter, we restrict our analysis to configuration (i) [Eq.~\eqref{eq:d_1}]; results for configuration (ii) [Eq.~\eqref{eq:d_2}] are provided in the End Matter. For $\theta = 0$, the model reduces to the antiferromagnetic Heisenberg limit, whose ground state is the well-known Neel order. In contrast, $\theta = \pi/2$ corresponds to the Heisenberg-Kitaev limit (with $J_{\rm K}/J_{\rm H}=-2$), extensively studied in the context of Kitaev materials, where the system stabilizes a collinear zigzag state.\cite{chaloupka2013zigzag,rau2014generic,rousochatzakis2015phase,rau2016spin,hermanns2018physics,3m4m-3v59,schaffer2012quantum} Accordingly, models described by Eq.~\eqref{eq:hubbard} are often termed Kitaev–Hubbard models in the literature. Related Kitaev-Hubbard models have been studied in condensed-matter and cold-atom contexts.~\cite{duan2003controlling,hassan2013quarter,faye2014topological,sato2021quantum,rachel18rpp116501} The key distinction to these works is that the spin-dependent hopping term in Eq.\,\eqref{eq:hubbard} is imaginary. Thus, the non-interacting part of the Hamiltonian preserves time-reversal symmetry, and generates the DM interactions in the strong-coupling limit.

\textbf{\textit{Classical spin model.---}}
To gain initial insight into the nature of the magnetic ground state, we first analyze the classical spin model, before proceeding to the quantum case. In the classical limit, quadratic spin Hamiltonians can be treated exactly within the Luttinger–Tisza (LT) framework, where the classical ground state is obtained by minimizing the Fourier-transformed interaction matrix.~\cite{luttinger1946theory,sklan2013nonplanar,sim2018discovery} When the lowest-energy LT eigenmodes satisfy the local hard-spin constraint $|\mathbf{S}_i| = 1$—either individually or through a linear combination of degenerate eigenmodes—the LT solution yields the exact classical ground state. In the entire parameter regime explored in this work, the LT solution satisfies the hard-spin constraint. Iterative minimization~\cite{sklan2013nonplanar} is employed in a complementary manner to resolve how the degenerate LT eigenmodes combine into a physical state that satisfies the hard-spin constraint. The resulting classical phase diagram of the model in Eq.~(\ref{eq:spin}) is shown in Fig.~\ref{fig_1}(a) (see Sec.~S1 of the Supplemental Material~\cite{supplemental}). For $0 < \theta < 0.85$, the system favors an incommensurate spiral state, which can be identified from the eigenvalue spectrum of the LT interaction matrix. Fig.~\ref{fig_1}(c) displays the LT spectrum for $\theta = \pi/5$, where the minimum of the lowest band lies between the high-symmetry points $\Gamma$ and M. The corresponding eigenmode at this wave vector satisfies the hard-spin constraint, indicating that the system stabilizes an incommensurate spiral state in this parameter regime.

For $0.85 \le \theta < \pi/2$, the minimum of the lowest LT band is located exactly at the M point, as shown in Fig.~\ref{fig_1}(d) for $\theta = \pi/3$, and the corresponding eigenmode satisfies the hard-spin constraint. In this regime, the system stabilizes a noncollinear zigzag state, which corresponds exactly to a \textit{$p$-wave magnet}. The relative canting angle $\phi$ between the spin-polarization axes $S^{\tilde z}_A$ and $S^{\tilde z}_B$ of the two crystallographic sublattices, shown in Fig.~\ref{fig_1}(b), quantifies the degree of noncollinearity: at $\phi = 0$, the system is in a collinear zigzag state, whereas $0 < \phi < \pi$ corresponds to the $p$-wave magnetic state; the limiting case $\phi = \pi$ corresponds to a collinear stripy state. The canting angle decreases continuously as $\theta$ approaches $\pi/2$, as shown in Fig.~\ref{fig_1}(a). In this phase, the magnetic unit cell is doubled relative to the crystallographic unit cell, as depicted in Fig.~\ref{fig_1}(b). The $p$-wave state is, however, energetically degenerate with noncoplanar states constructed as a superposition of three symmetry-related degenerate eigenmodes whose ordering wave vectors lie at the M points connected by the $C_{3z}$ rotation (see Sec.~S1 of the Supplemental Material~\cite{supplemental}). In Figs.~\ref{fig_1}(e,f), we present the nearest-neighbor spin correlators $\langle S^{\tilde z}_{i} S^{\tilde z}_{j} \rangle$ and $\langle S^{\tilde z}_{i} S^{\tilde x}_{j} \rangle$ for the $p$-wave magnet. Like $S^{\tilde z}_i$, the spin component $S^{\tilde x}_{i\in A,B}$ along the local $\tilde{x}$ axis is defined separately on sublattices $A$ and $B$. The correlation pattern indicates that spins are collinear within their respective local frames but rotate relative to each other in the global frame, giving rise to a globally noncollinear yet strictly coplanar configuration. In contrast, the competing multi-$Q$ noncoplanar states display finite out-of-plane correlations, $\langle S^{\tilde z}_{i} S^{\tilde x}_{j} \rangle$, and exhibit a qualitatively distinct correlation pattern (see Sec.~S1 of the Supplemental Material for details~\cite{supplemental}).

It is expected that quantum order-by-disorder effects favor the single-$Q$ coplanar state over the multi-$Q$ noncoplanar states.~\cite{henley1989ordering,PhysRevB.46.11137} Below, we move on to the quantum model and investigate whether a $p$-wave magnet is indeed selected by quantum fluctuations.

\begin{figure}[]
	\includegraphics[width = 1.0\columnwidth]{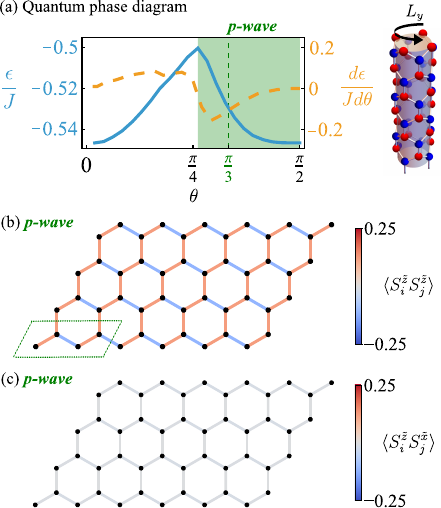}
	\caption{(a) Quantum phase diagram of the spin-1/2 model in Eq.~(\ref{eq:spin}) for $\hat{\bs d}_\gamma = \hat{\gamma}$. The simulations are performed on an infinite cylinder with $L_x \times L_y \times 2 = 6 \times 4 \times 2$ sites per unit cell, where $L_x$ is repeated along the infinite direction and $L_y$ is the periodic circumference. Quantum fluctuations select the $p$-wave magnet over the noncoplanar states that are energetically degenerate at the classical level. (b,c) Nearest-neighbor spin correlators $\langle S^{\tilde z}_{i} S^{\tilde z}_{j} \rangle$ and $\langle S^{\tilde z}_{i} S^{\tilde x}_{j} \rangle$ for the $p$-wave magnet at $\theta = \pi/3$. The magnetic unit cell is indicated by the dashed green outline in (b).}
	\label{fig_2}
\end{figure}

\textbf{\textit{Quantum spin model.---}}
To clarify which features of the classical phase diagram survive in the quantum regime, we perform iDMRG~\cite{mcculloch2008infinite,schollwock2011density} calculations on the spin-1/2 model. The simulations employ a unit cell containing 48 sites and bond dimensions up to 2000, providing a reliable description of the phases found at the classical level, except for the incommensurate spiral states (see Sec.~S2 of the Supplemental Material for details~\cite{supplemental}). The results are presented in Fig.~\ref{fig_2}(a). A first-order phase transition is indicated by a kink in the ground-state energy per site $\epsilon$, where the first derivative $\partial \epsilon / \partial \theta$ changes sign at $\theta \simeq 0.85$. To characterize the state stabilized in $0.85 \leq \theta < \pi/2$, we compute the nearest-neighbor spin correlators $\langle S^{\tilde z}_{i} S^{\tilde z}_{j} \rangle$ and $\langle S^{\tilde z}_{i} S^{\tilde x}_{j} \rangle$, as shown in Figs.~\ref{fig_2}(b,c). The correlation patterns exhibit the same structure as in the classical case shown in Figs.~\ref{fig_1}(e,f): spins remain collinear within their respective local reference frames but form a globally noncollinear yet coplanar configuration. The resulting magnetic structure has a four-site magnetic unit cell, as shown in Fig.~\ref{fig_2}(b), consistent with the $p$-wave magnet. Thus, quantum fluctuations lift the classical degeneracy and select the $p$-wave magnet as the ground state.

\textbf{\textit{Band structure and Edelstein effect.---}}
The $p$-wave magnet possesses distinctive symmetry properties. It breaks spatial inversion $P$ as well as the combined $P\mathcal{T}$ symmetry, where $\mathcal{T}$ denotes time reversal. At the same time, it preserves the composite symmetry $\mathcal{T} T(\bs{t}_1)$, where $\bs{t}_1$ is one of the lattice translation vectors and $T(\bs{t}_1)$ is the corresponding translation operator. This symmetry preserves time-reversal symmetry in momentum space, enforcing the relation $E(\bs{k}, \bs{S}) = E(-\bs{k}, -\bs{S})$. In addition, the system retains a spin symmetry $[C_2^{\perp}||\bs{t}_1]$, where $C_2^{\perp}$ denotes a 180$^\circ$ spin rotation about the axis perpendicular to the spin plane. The corresponding symmetry $[C_2^{\perp}||E]$ in momentum space enforces that only the spin component along the $C_2^{\perp}$ rotation axis remains finite, resulting in a spin polarization strictly aligned with this axis. Together, these two symmetries constrain the spin polarization in momentum space to satisfy
\bea
E(\bs{k}, S^{\perp}) = E(-\bs{k}, -S^{\perp}).
\label{eq:momentum_spin}
\eea

\begin{figure}[]
	\includegraphics[width = 1.0\columnwidth]{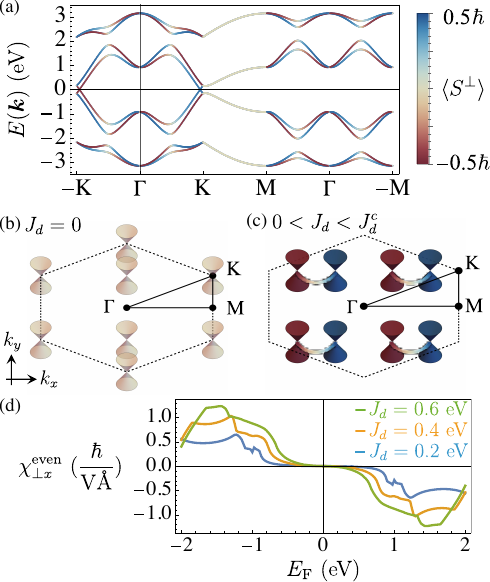}
	\caption{(color online) (a) Band spectrum of the model in Eq.~(\ref{eq:kondo}) with parameters $t = 1$ eV and $J_d = 0.6$ eV. The bands exhibit $p$-wave spin polarization in momentum space, consistent with Eq.~(\ref{eq:momentum_spin}). Spin degeneracy is restored along the high-symmetry $\Gamma$–$M$ line at the Brillouin-zone boundary. (b) At $J_d = 0$, four spin-degenerate Dirac cones appear at the Fermi level within the first Brillouin zone of the crystallographic unit cell. (c) For $0 < J_d < J_d^c$, eight spin-polarized Dirac cones appear at the Fermi level in the first Brillouin zone of the crystallographic unit cell. At the critical coupling $J_d = J_d^c$, Dirac cones annihilate in pairs and a gap opens. Pairs of Dirac cones are connected by guiding lines for visualization. (d) Calculated $\chi^{\text{even}}_{\perp x}$ for $t = 1$ eV and $\Gamma = 0.01$ eV with $J_d = 0.2$, $0.4$, and $0.6$ eV. The magnitude of $\chi^{\text{even}}_{\perp x}$ increases with $J_d$.}
	\label{fig_3}
\end{figure}

We introduce a minimal tight-binding model to illustrate the electronic properties of the $p$-wave magnet on the honeycomb lattice. The model is given by
\bea
H_{\text{p}} = t \sum_{\langle i j \rangle} c_{i}^\dagger c_{j}
+ J_d \sum_i c_{i}^\dagger [\hat{\bs{m}}_i^{\mathrm{p}} \!\cdot\! \bs{\sigma}] c_{i},
\label{eq:kondo}
\eea
where $\hat{\bs{m}}_i^{\mathrm{p}}$ denotes the site-dependent local spin polarization of the $p$-wave magnet shown in Fig.~\ref{fig_1}(b), and $J_d$ quantifies the coupling between the itinerant electrons and the localized moments. In the following, we take $\hat{\bs{m}}_i^{\mathrm{p}}$ from the $p$-wave magnetic configuration stabilized at $\theta=\pi/3$ in Eq.~(\ref{eq:spin}), which corresponds to the canting angle $\phi=1.22$. The electronic band structure of Eq.~(\ref{eq:kondo}) is shown in Fig.~\ref{fig_3}(a), plotted within the first Brillouin zone of the crystallographic unit cell (the number of bands is doubled due to zone folding). The resulting band dispersion clearly exhibits $p$-wave spin splitting, satisfying the relation given in Eq.~(\ref{eq:momentum_spin}). Such splitting is absent in the collinear zigzag and stripy states, corresponding to $\phi=0$ and $\phi=\pi$, respectively (see Sec.~S3 of the Supplemental Material for details~\cite{supplemental}). Moreover, we find that the spin-degenerate Dirac cones present at $J_d = 0$ [Fig.~\ref{fig_3}(b)] split into pairs of spin-polarized Dirac cones once $J_d > 0$, as shown in Fig.~\ref{fig_3}(c). These Dirac cones remain stable until $J_d$ reaches a critical value $J_d^c$, where Dirac cones with opposite spin polarization annihilate in pairs. On the other hand, the Dirac cones are immediately gapped once sublattice symmetry is broken—for instance, by introducing a sublattice-dependent onsite potential.

The odd-parity spin–momentum locking enables an electric current to induce a net spin polarization—a magnetoelectric response known as the Edelstein effect~\cite{freimuth2014spin,li2015intraband,PhysRevLett.119.196801,manchon2019current}. To evaluate this effect, we employ the Kubo formalism within the constant relaxation time approximation. We focus on the component of the response that is linear in the applied electric field, written as $\delta S_\alpha= \chi_{\alpha \beta} E_\beta$, where $\delta S_\alpha$ is the induced spin polarization, $E_\beta$ is the electric field, and $\chi_{\alpha \beta}$ is the response tensor. Time-reversal even and odd components of $\chi_{\alpha \beta}$ are derived in Ref.~\onlinecite{freimuth2014spin,li2015intraband}:
\small
\bea
\nonumber
\chi^{\text{even}}_{\alpha \beta} &=& \frac{-e \hbar}{\pi}
\sum_{m,n,\bs k}
\frac{\mathrm{Re}\!\left[
	\langle \psi_{\bs k,m} | \hat S_\alpha | \psi_{\bs k,n} \rangle
	\langle \psi_{\bs k,n} | \hat v_\beta | \psi_{\bs k,m} \rangle
	\right] \Gamma^2}
{[(E_\mathrm{F} - E_{\bs k,m})^2 + \Gamma^2]
	[(E_\mathrm{F} - E_{\bs k,n})^2 + \Gamma^2]},
\\
\label{eq:response_even}
\eea
\bea
\nonumber
\chi^{\text{odd}}_{\alpha \beta} &=&
2 e \hbar
\sum_{\substack{m: \mathrm{occ.}\\ n: \mathrm{unocc.}},\,\bs k}
\mathrm{Im}\!\left[
\langle \psi_{\bs k,n} | \hat S_\alpha | \psi_{\bs k,m} \rangle
\langle \psi_{\bs k,m} | \hat v_\beta | \psi_{\bs k,n} \rangle
\right]
\\
&\times&
\frac{\Gamma^2 - (E_{\bs k,m} - E_{\bs k,n})^2}
{[(E_{\bs k,m} - E_{\bs k,n})^2 + \Gamma^2]^2}.
\label{eq:response_odd}
\eea
\normalsize
Here, $\ket{\psi_{\bs k, n}}$ denotes the Bloch eigenstate of band $n$ with wave vector $\bs k$, whose energy is $E_{\bs k,n}$, and $E_{\mathrm{F}}$ is the Fermi level. The quantity $e>0$ is the elementary charge, $\hat{S}_\alpha$ is the spin operator, and $\hat{v}_\beta = \frac{1}{\hbar}\partial H_\text{p}(\bs k)/\partial k_\beta$ is the group-velocity operator. The parameter $\Gamma$ characterizes disorder broadening and is related to the relaxation time by $\tau = \hbar/(2\Gamma)$.
As $\chi^{\mathrm{odd}}$ represents a Fermi-surface property that is odd under time-reversal symmetry, it is allowed only in systems that explicitly break the time-reversal symmetry, which include ferromagnets and certain non-collinear antiferromagnets~\cite{vzelezny2017spin,manchon2019current,gurung2021transport,gonzalez2024non}. In our case of the $p$-wave state, this quantity vanishes, as also confirmed by our numerical calculations, in agreement with the result in Ref.~\onlinecite{chakraborty2025highly}. 

For the model in Eq.~(\ref{eq:kondo}), $\chi^{\text{even}}_{\perp x}$ is the only nonvanishing component of the Edelstein response tensor. This corresponds to a current-induced spin accumulation oriented along the band spin-polarization axis when an electric field is applied along the $x$ direction. In Fig.~\ref{fig_3}(d), we plot $\chi^{\text{even}}_{\perp x}$ as a function of the Fermi level $E_{\text{F}}$ for $t = 1$ eV, $\Gamma = 0.01$ eV, and several values of $J_d$. As $J_d$ increases, Dirac cones with opposite spin polarization are displaced further in opposite $k_x$ directions, resulting in an enhanced response magnitude. The particle–hole symmetry observed in the response is consistent with the band structure shown in Fig.~\ref{fig_3}(a). To indicate the typical scale of the response, we note that comparable susceptibility values have been reported in other magnetic systems. For example, a $p$-wave magnet on a square lattice studied in Ref.~\onlinecite{chakraborty2025highly} yields $\chi^{\text{even}} \simeq 2\hbar/\mathrm{V\AA}$ for $\Gamma = 0.01$eV. Within our minimal model description, the Edelstein response lies within a similar range.

\textbf{\textit{Summary and outlook.---}} In summary, we have identified Hubbard models that realizes a $p$-wave magnet with a commensurate ordering wave vector as its ground state. In the strong-coupling limit $U/t \to \infty$, the corresponding low-energy spin Hamiltonian stabilizes a $p$-wave magnetic state that is energetically degenerate with a competing noncoplanar state at the classical level. In the quantum limit $S = 1/2$, this degeneracy is lifted by quantum fluctuations, which select the $p$-wave magnet. We then introduce a minimal model in which itinerant electrons couple to this commensurate noncollinear magnetic texture and analyze the resulting band structure. The model exhibits a defining feature of $p$-wave magnetism: an odd-parity spin-polarized band dispersion across the Brillouin zone. Furthermore, we compute the Edelstein response and find a finite symmetry-allowed component of the response tensor $\chi^{\text{even}}_{\perp x}$, demonstrating current-induced spin accumulation along the band spin-polarization axis.

Although our main analysis focuses on case (i) defined by Eq.~(\ref{eq:d_1}), case (ii) specified in Eq.~(\ref{eq:d_2}) is directly relevant  to the honeycomb magnet Ni$_2$Mo$_3$O$_8$, which hosts effective spin-1 moments on Ni$^{2+}$ ions. In this compound, neighboring Ni ions occupy inequivalent tetrahedral and octahedral oxygen environments, breaking inversion symmetry on each nearest-neighbor bond.\cite{morey2019n, PhysRevResearch.5.033099} This local asymmetry permits finite DM interactions with DM vectors that correspond to the pattern in Eq.~(\ref{eq:d_2}).\cite{gao2023diffusive} Such DM interactions can stabilize the commensurate $p$-wave magnet, namely the noncollinear zigzag state that has been experimentally observed in Ni$_2$Mo$_3$O$_8$ by neutron scattering measurements~\cite{gao2023diffusive}. See End Matter for details.

While we have focused on the strong-coupling limit in this work, an important open question is whether the $p$-wave magnet also emerges in the weak-coupling regime. In addition, now that effective spin models realizing the $p$-wave magnet have been established, a natural next step is to investigate its magnon excitation spectrum and associated dynamical properties. Another promising direction is to examine whether lattice distortions can generate finite DM interactions in Kitaev candidate materials by lifting local inversion symmetry on the relevant bonds. This could provide a route to the additional anisotropic exchanges needed to stabilize commensurate $p$-wave magnetism. A representative example is the Kitaev material $\beta$-Li$_2$IrO$_3$, where the presence of weak DM interactions was predicted~\cite{yadav2018strain}.

\textbf{\textit{Acknowledgments.---}}  
We thank Matthew Bunney, Themba Hodge, and Matthias Vojta for insightful discussions related to this work. G.B.S. thanks Johannes Knolle, Moon Jip Park, and SungBin Lee for hosting during the initial stage of this work. This work was supported by the Australian Research Council (ARC) through Grant No.\ DP240100168. This research was undertaken using resources from the National Computational Infrastructure (NCI Australia), an NCRIS enabled capability supported by the Australian Government. G.B.S. acknowledges support from the National Research Foundation of Korea (NRF) through Grant No. RS-2024-00453943. Tensor network calculations were performed using the TeNPy Library~\cite{10.21468/SciPostPhysLectNotes.5,10.21468/SciPostPhysCodeb.41}.

\textbf{\textit{Data availability.---}}
Raw data and simulation codes are available in Zenodo upon reasonable request~\cite{data_zenodo}.

\begin{widetext}
\section*{End Matter}
\end{widetext}

\textit{Case (ii): Classical and spin-$1/2$ phase diagrams}---
Here, we focus on case (ii), specified by the pattern in Eq.~(\ref{eq:d_2}). The resulting classical phase diagram of Eq.~(\ref{eq:spin}) is shown in Fig.~\ref{fig_4}(a) (see Sec.~S4 of the Supplemental Material for details~\cite{supplemental}). For small angles $0 < \theta < 0.67$, the ground state is an incommensurate spiral. Increasing $\theta$ drives a transition into the $p$-wave magnet, which is stabilized for $0.67 \le \theta < 1.25$ and is characterized by the ordering wave vector $\mathbf{Q} = \mathrm{M}$, as indicated by the LT spectrum in Fig.~\ref{fig_4}(b). Within this regime, the $p$-wave magnet is classically degenerate with noncoplanar states. Upon further increasing $\theta$, the system enters a coplanar $120^\circ$ state for $1.25 \le \theta < \pi/2$, with ordering wave vector $\mathbf{Q} = \mathrm{K}$, as shown in Fig.~\ref{fig_4}(c). This state may be viewed as a coplanar spiral with spins confined to a plane perpendicular to the [001] axis, where spins on the same sublattice are separated by relative angles of $\pm 2\pi/3$, while the relative orientations between different sublattices remain unconstrained. The spin group symmetry of such magnetic order can give rise to an $f$-wave spin splitting in momentum space~\cite{luo2025spin}.

\begin{figure}[]
	\includegraphics[width = 1.0\columnwidth]{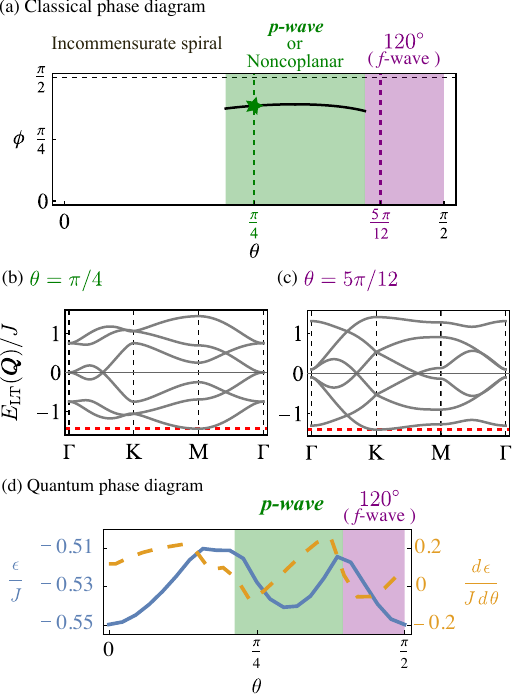}
	\caption{(a) Classical phase diagram of the model given in Eq.~(\ref{eq:spin}) for the $\hat{\bs d}_{\gamma}$ specified in Eq.~(\ref{eq:d_2}). For $0.67 \le \theta < 1.25$, the model stabilizes the $p$-wave magnetic state and a competing noncoplanar state, which are energetically degenerate. The canting angle $\phi$ is defined for the $p$-wave magnet and quantifies the noncollinearity between the two sublattices. (b,c) Eigenvalue spectra of the LT interaction matrix for $\theta = \pi/4$ and $5\pi/12$. The minimum of the lowest-energy band is located at the M point for the former and at the K point for the latter. (d) Quantum phase diagram of the corresponding spin-1/2 model.}
	\label{fig_4}
\end{figure}

To examine the stability of these states in the quantum limit, we study the spin-1/2 model using iDMRG on an infinite cylinder with $L_x \times L_y \times 2 = 6 \times 3 \times 2$ sites per unit cell (see Sec.~S5 of the Supplemental Material~\cite{supplemental}). This geometry is commensurate with the magnetic unit cells of both the $p$-wave magnet and the coplanar $120^\circ$ state, while incommensurate spiral states are suppressed by the finiteness of the unit-cell size. As shown in Fig.~\ref{fig_4}(d), the ground-state energy per site $\epsilon$ develops kinks at $\theta \simeq 0.67$ and $1.25$. Correspondingly, the first derivative $d\epsilon/d\theta$ exhibits two sign changes at these values, signaling first-order phase transitions consistent with the classical phase diagram. Similar to case (i), quantum fluctuations lift the classical degeneracy and select the $p$-wave magnet over competing noncoplanar states for $0.67 \le \theta < 1.25$.

\begin{figure}[]
	\includegraphics[width = 0.9\columnwidth]{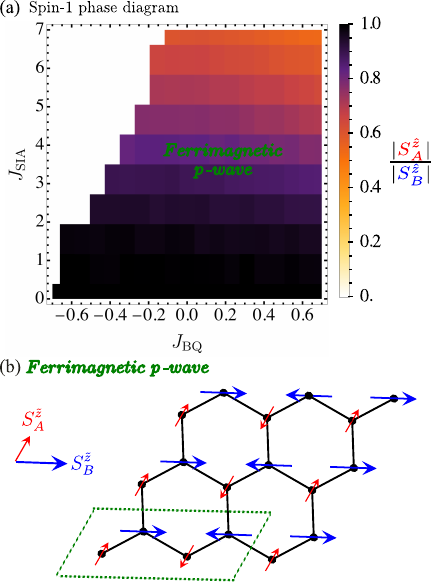}
	\caption{(a) iDMRG phase diagram of the spin-1 model in Eq.~(\ref{eq:spin_1}) with $\hat{\bs d}_{\gamma}$ specified in Eq.~(\ref{eq:d_2}) at fixed $\theta=0.95$, shown as a function of $J_{\mathrm{BQ}}$ and $J_{\mathrm{SIA}}$. A finite sublattice-dependent SIA stabilizes a ferrimagnetic $p$-wave state over a broad parameter regime. (b) Spin configuration of the ferrimagnetic $p$-wave state. The local magnetic moments on sublattices $A$ and $B$ have different magnitudes. The magnetic unit cell, indicated by the dashed green outline, is twice as large as the crystallographic unit cell.}
	\label{fig_5}
\end{figure}

\textit{Case (ii): Spin-1 extension relevant to Ni$_2$Mo$_3$O$_8$}---
Here, we focus on the spin-1 extension of case (ii), motivated by the honeycomb magnet Ni$_2$Mo$_3$O$_8$~\cite{morey2019n, PhysRevResearch.5.033099,gao2023diffusive}. In this material, the magnetically active Ni$^{2+}$ ions occupy inequivalent tetrahedral and octahedral oxygen environments, which break bond-centered inversion symmetry and thereby allow finite DM interactions. Since the Ni$^{2+}$ ions carry local spin-1 moments, the effective model generally contains additional terms beyond Eq.~(\ref{eq:spin}). To capture these effects, we consider the following spin-1 generalization:
\bea
\nonumber
H_s &=& \sum_{\langle ij \rangle \in \gamma} \big[\cos{2\theta}\, \bs{S}_i \!\cdot\! \bs{S}_j
+ \sin{2\theta}\, \hat{\bs d}_{\gamma} \!\cdot\! (\bs{S}_i \!\times\! \bs{S}_j)
\\
\nonumber
&+& (1 - \cos{2\theta}) (\hat{\bs d}_{\gamma} \!\cdot\! \bs{S}_i)(\hat{\bs d}_{\gamma} \!\cdot\! \bs{S}_j)
\big]
+ J_{\text{BQ}} \sum_{\langle ij \rangle} (\bs{S}_i \cdot \bs{S}_j)^2
\\
&+& J_{\text{SIA}} \big[ \sum_{i \in A} (S_i^z)^2 + \frac{1}{16}\sum_{i \in B} (S_i^z)^2 \big],
\label{eq:spin_1}
\eea
where $J_{\text{BQ}}$ denotes the strength of the biquadratic (BQ) exchange and $J_{\text{SIA}}$ quantifies the single-ion anisotropy (SIA). Apart from these terms, Eq.~(\ref{eq:spin_1}) has the same form as the model introduced in Eq.~(\ref{eq:spin}). The last term represents sublattice-dependent SIA. In particular, we take the anisotropy on sublattice $B$ to be $1/16$ of that on sublattice $A$, following the value used in Ref.~\onlinecite{gao2023diffusive}. This choice reflects the strong sublattice asymmetry of the local crystal-field anisotropy in Ni$_2$Mo$_3$O$_8$.

Using iDMRG at a fixed value of $\theta=0.95$, we obtain the phase diagram shown in Fig.~\ref{fig_5}(a). The results demonstrate that a finite sublattice-dependent SIA renders the $p$-wave state \textit{ferrimagnetic}. The corresponding magnetic structure is shown in Fig.~\ref{fig_5}(b), where the local magnetic moments on sublattices $A$ and $B$ have unequal magnitudes. Despite this imbalance, the system still preserves the composite symmetry $\mathcal{T} T(\bs{t}_1)$ and the spin symmetry $[C_2^{\perp} || \bs{t}_1]$. Our iDMRG results show that the $p$-wave state is stable over a broad parameter regime in $J_{\mathrm{SIA}}$ and $J_{\mathrm{BQ}}$ [Fig.~\ref{fig_5}(a)].

\newpage

\renewcommand{\thefigure}{S\arabic{figure}}
\setcounter{figure}{0}
\renewcommand{\theequation}{S\arabic{equation}}
\setcounter{equation}{0}
\renewcommand{\thetable}{S\arabic{table}}
\setcounter{table}{0}

\begin{widetext}
\section{Supplementary Material}

Here, we present the detailed analysis underlying the results shown in the main text. Specifically, we provide the explicit form of the Luttinger–Tisza (LT) interaction matrix, the LT eigenvalue spectra across the full parameter range considered in the main text, the details of the iterative minimization procedure, and the details of the matrix product state simulations.
\subsection{S1. Classical spin model for case (i)}
\subsubsection{LT method}
\begin{figure}[h!]
	\includegraphics[width = 0.28\columnwidth]{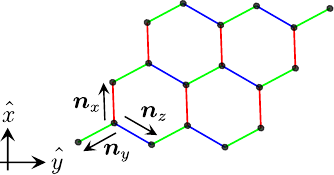}
	\caption{2D honeycomb lattice, with three types of nearest-neighbor bonds, labeled by $\gamma\!=\!x$, $y$ or $z$. The axes $\hat x$ and $\hat y$ define the plane of the lattice,  $\bs{n}_x=a(\frac{1}{\sqrt 3}\hat y)$, $\bs{n}_y=a(-\frac{1}{3} \hat x - \frac{1}{2\sqrt3} \hat y)$, $\bs{n}_z=a(\frac{1}{3} \hat x - \frac{1}{2\sqrt3} \hat y)$, and $a$ is the lattice constant.}
	\label{s_fig_1}
\end{figure}
Fig.~\ref{s_fig_1} shows the honeycomb lattice and our convention for the nearest-neighbor bond vectors $\bs n_x$, $\bs n_y$, and $\bs n_z$. For case (i), the spin Hamiltonian is written as
\bea
H_s &=& J \sum_{\langle ij \rangle \in \gamma} \big[\cos{2\theta}\, \bs{S}_i \cdot \bs{S}_j
+ \sin{2\theta}\, \hat{\bs d}_{\gamma} \cdot (\bs{S}_i \!\times\! \bs{S}_j)+ (1 - \cos{2\theta}) (\hat{\bs d}_{\gamma} \cdot \bs{S}_i)(\hat{\bs d}_{\gamma} \cdot \bs{S}_j)
\big]
\label{s_eq:spin}
\eea
where
\bea
\hat{\bs d}_{x} = (1,0,0), \hat{\bs d}_{y} = (0,1,0), \hat{\bs d}_{z} = (0,0,1).
\eea 
Upon Fourier transforming the spin operators, Eq.~(\ref{s_eq:spin}) can be written as
\bea
H_s=\sum_{\bs{Q}} \left(\bs{S}_{-\bs Q,A}^T,\bs{S}_{-\bs{Q},B}^T\right) \cdot 
H_\text{LT}(\bs Q ) \cdot 
\left(\begin{array}{c}
	\bs{S}_{\bs{Q},A} \\
	\bs{S}_{\bs{Q},B}
\end{array}\right)
\label{s_eq:H_LT}
\eea
where $\bs S_{\bs Q, \nu}=\frac{1}{\sqrt{N/2}}\sum_{i \in \nu} \bs{S}_i e^{-i \bs Q \cdot \bs r_i}$ and $\bs{S}_{\bs{Q},\nu}\!=\!(S_{\bs{Q},\nu}^x,S_{\bs{Q},\nu}^y,S_{\bs{Q},\nu}^z)^T$.~\cite{luttinger1946theory,sklan2013nonplanar} The 6$\times$6 LT interaction matrix $H_\text{LT}(\bs Q )$ is given by:
\bea
\nonumber
H_\text{LT}(\bs Q) = \frac{J}{2}\left(
\begin{array}{cc}
	0 & h(\bs Q) \\
	h^T(-\bs Q) & 0
\end{array}
\right),
\eea
\bea
h(\bs Q)=
\left(
\begin{array}{ccc}
	e^{i\bs n_x\bs Q}
	\!+\!e^{i\bs n_y\bs Q}\,J_\text{H}
	\!+\!e^{i\bs n_z\bs Q}\,J_\text{H}
	&
	-\,J_\text{D}\,e^{i\bs n_z\bs Q}
	&
	\;\;J_\text{D}\,e^{i\bs n_y\bs Q}
	\\
	J_\text{D}\,e^{i\bs n_z\bs Q}
	&
	e^{i\bs n_y\bs Q}
	\!+\!e^{i\bs n_x\bs Q}\,J_\text{H}
	\!+\!e^{i\bs n_z\bs Q}\,J_\text{H}
	&
	-\,J_\text{D}\,e^{i\bs n_x\bs Q}
	\\
	-\,J_\text{D}\,e^{i\bs n_y\bs Q}
	&
	J_\text{D}\,e^{i\bs n_x\bs Q}
	&
	e^{i\bs n_z\bs Q}
	\!+\!e^{i\bs n_x\bs Q}\,J_\text{H}
	\!+\!e^{i\bs n_y\bs Q}\,J_\text{H}
\end{array}
\right)
\label{s_eq:h_Q}
\eea
where $ J_\text{H}=\cos 2\theta$ and $J_\text{D}=\sin 2\theta$. 

The eigenspectrum of the LT matrix, $H_\text{LT}(\bs Q)$, is shown in Fig.~\ref{s_fig_2} for several values of $\theta$. At $\theta=0$, the minimum lies at $\Gamma$, yielding a collinear antiferromagnet. For $0 < \theta < 0.85$, the minimum occurs at an incommensurate wave vector and the corresponding eigenmode satisfies the hard-spin constraint. In this range the system stabilizes an incommensurate spiral state. For $0.85 \le \theta < \pi/2$, the minimum of the lowest LT band moves to the M manifold: $M = \frac{1}{a}\left(\pi,-\frac{\pi}{\sqrt{3}}\right)$, $C_{3z} M = \frac{1}{a}\left(0,\frac{2\pi}{\sqrt{3}}\right)$, and $C_{3z}^2 M = \frac{1}{a}\left(\pi,\frac{\pi}{\sqrt{3}}\right)$, and each corresponding eigenmode individually satisfies the hard-spin constraint. In this regime, the noncollinear zigzag state ($p$-wave magnet) as a single-$\bs Q$ state and the noncoplanar states as linear combinations of multi-$\bs Q$ states are energetically degenerate. Likewise, at $\theta=\pi/2$ the system supports both the collinear zigzag single-$\bs Q$ state and the noncoplanar multi-$\bs Q$ states, which are energetically degenerate at the classical level.
\begin{figure}[]
	\includegraphics[width = 0.68\columnwidth]{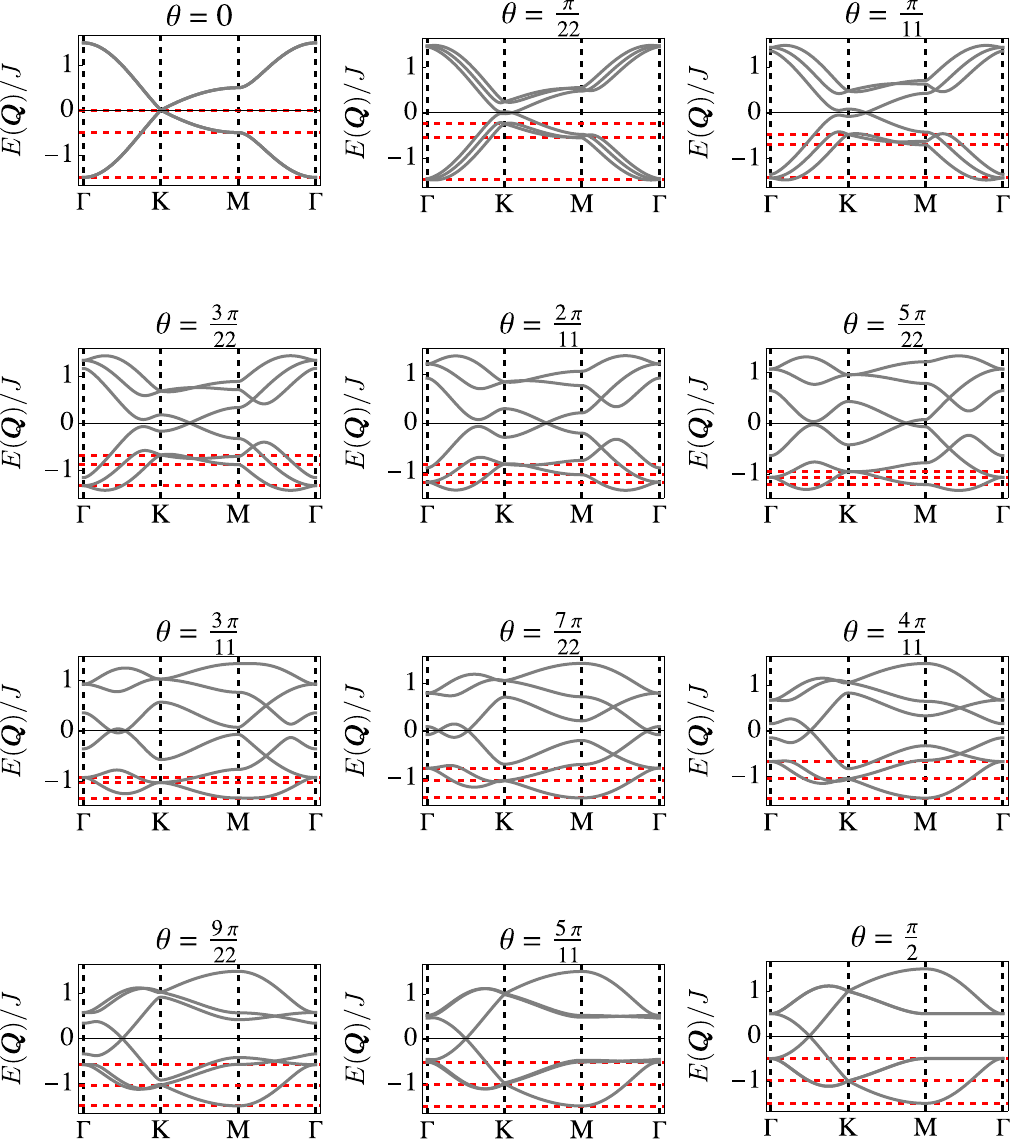}
	\caption{Eigenspectrum of the LT matrix for case (i). For reference, the lowest eigenvalue at the high-symmetry points $\Gamma$, K, and M is indicated by three red dashed horizontal lines in each panel.}
	\label{s_fig_2}
\end{figure}

\subsubsection{Iterative minimization}
To select a physical spin configuration within a classically degenerate LT ground state manifold, we perform iterative minimization~\cite{sklan2013nonplanar} on finite periodic clusters. We work on an $L_x\times L_y$ array of honeycomb unit cells with two sublattices, so the total number of spins is $N=2L_xL_y$. Nearest-neighbor bonds and their $3\times3$ interaction matrices define a symmetric $3N\times3N$ interaction matrix $\overset{\leftrightarrow}{\bs H}_{\text{IM}}$ acting on the spin vector $\bs S=(S_1^x,S_1^y,S_1^z;\ldots;S_N^x,S_N^y,S_N^z)^T$. We initialize $\{\bs S_i^{(0)}\}$ as independent random unit vectors on $S^2$. Each iteration updates all spins by
\bea
\nonumber
&&\bs H^{(t)}= \overset{\leftrightarrow}{\bs H}_{\text{IM}} \bs S^{(t)},\\
\nonumber
&&\widetilde{\bs S}^{(t+1)} = \bs S^{(t)} - \eta\,\bs H^{(t)},\\
&&\bs S^{(t+1)}_i = \frac{\widetilde{\bs S}^{(t+1)}_i}{\big|\widetilde{\bs S}^{(t+1)}_i\big|} \quad \text{for all } i=1,\dots,N,
\eea
i.e., a global gradient step with step size $\eta$ followed by sitewise projection back to the unit sphere. Convergence is monitored by
\bea
\varepsilon^{(t)}=\frac{\big\|\bs S^{(t+1)}-\bs S^{(t)}\big\|}{N},
\eea
and the run is terminated once $\varepsilon^{(t)}<\varepsilon_\star$. In practice we used $\eta=0.02$ and $\varepsilon_\star=10^{-7}$. The classical energy per site is
\bea
E/N=\frac{\bs S^T \overset{\leftrightarrow}{\bs H}_{\text{IM}} \bs S }{N}.
\eea
We use periodic clusters up to $L_x=L_y=6$ (i.e. $L_x \times L_y \times 2=72$ spins), which is sufficient to capture all commensurate single-$\bs Q$ and multi-$\bs Q$ states. The incommensurate regime is not faithfully represented on such finite clusters. Apart from the incommensurate regime, the energy per site obtained from iterative minimization agrees exactly with the LT energy $E_{\text{min}}(\bs Q)$.

In Figs.~\ref{s_fig_3}(a,b) we show the common-origin plot~\cite{sklan2013nonplanar} of the $p$-wave magnet and one representative noncoplanar state that is classically degenerate in energy at $\theta=\pi/3$, obtained from iterative minimization. The spin configuration of the $p$-wave magnet is noncollinear but remains coplanar. For the noncoplanar state, we further compute the nearest-neighbor spin correlators $\langle S^{\tilde z}_i S^{\tilde z}_j\rangle$ and $\langle S^{\tilde z}_i S^{\tilde x}_j\rangle$ following the convention used in the main text, as shown in Figs.~\ref{s_fig_3}(c,d). $S^{\tilde z}_{i\in A,B}$ is defined along the polarization direction of the first spin in the unit cell separately for sublattice A and B, while $S^{\tilde x}_{i\in A,B}$ denotes the spin component along the corresponding local $\tilde x$ axis. The finite value of $\langle S^{\tilde z}_i S^{\tilde x}_j\rangle$ clearly reflects noncoplanarity in the global frame, in contrast to the $p$-wave magnet, where this mixed component vanishes.

\begin{figure}[]
	\includegraphics[width = 0.85\columnwidth]{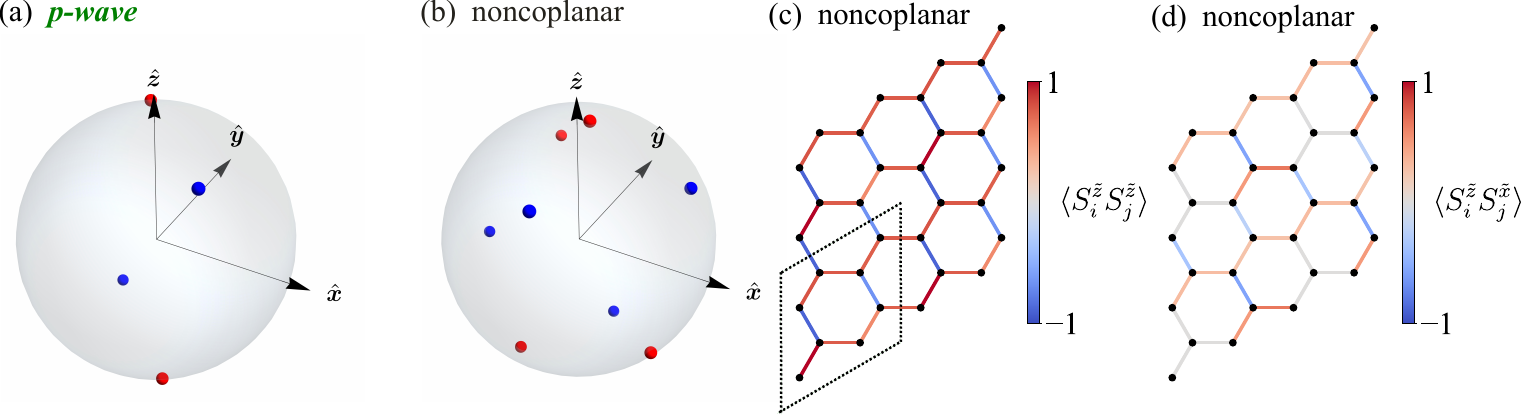}
	\caption{(a,b) Common-origin spin configurations of the $p$-wave magnet and one representative noncoplanar state that is classically degenerate in energy. Blue and red points denote sublattices A and B. (c,d) Nearest-neighbor spin correlators, $\langle S^{\tilde z}_i S^{\tilde z}_j\rangle$ and $\langle S^{\tilde z}_i S^{\tilde x}_j\rangle$, for the noncoplanar state. The magnetic unit cell contains eight sites, indicated by the dashed outline in (c).}
	\label{s_fig_3}
\end{figure}

\subsection{S2. Quantum spin model for case (i)}
We perform matrix-product-state calculations using the infinite density matrix renormalization group (iDMRG) for the quantum spin-1/2 model in Eq.~(\ref{s_eq:spin}).~\cite{mcculloch2008infinite,schollwock2011density} The calculations are carried out on an infinite cylinder where $L_x$ is repeated along the infinite direction and $L_y$ forms the periodic circumference. We use a unit cell of size $L_x \times L_y \times 2 = 6 \times 4 \times 2$ sites and keep bond dimensions up to $\chi=2000$. This setup provides a reliable description of the phases identified at the classical level, except for incommensurate spiral states.

To examine the convergence of the ground state obtained by iDMRG, we analyze the scaling of the energy per site as a function of bond dimension $\chi$. For $\theta=\pi/3$, the system stabilizes the $p$-wave magnet, as established in the main text. Fig.~\ref{s_fig_4} displays the corresponding convergence behavior: the energy decreases linearly with $1/\chi$, demonstrating reliable convergence.
\begin{figure}[]
	\includegraphics[width = 0.35\columnwidth]{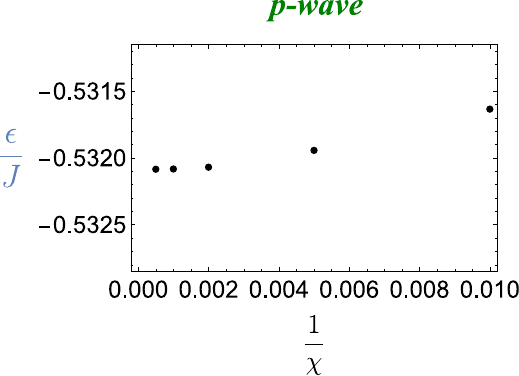}
	\caption{Bond dimension scaling of the ground–state energy at $\theta=\pi/3$.  The energy per site decreases approximately linearly with $1/\chi$, indicating that the iDMRG results are well converged.}
	\label{s_fig_4}
\end{figure}

\subsection{S3. Spin configurations and their corresponding band splitting}
In Fig.~\ref{s_fig_5}(a)-(c), we show the spin configurations of the collinear zigzag state, the $p$-wave state, and the collinear stripy state. These states are distinguished by the relative canting angle $\phi$ between the spin-polarization axes $S_A^{\tilde z}$ and $S_B^{\tilde z}$ on the two crystallographic sublattices: $\phi=0$ corresponds to the collinear zigzag state, $0<\phi<\pi$ to the $p$-wave state, and $\phi=\pi$ to the collinear stripy state. All three states preserve the composite symmetry $\mathcal{T} T(\bs{t}_1)$ and the spin symmetry $[C_2^{\perp} || \bs{t}_1]$.

\begin{figure}[]
	\includegraphics[width = 0.9\columnwidth]{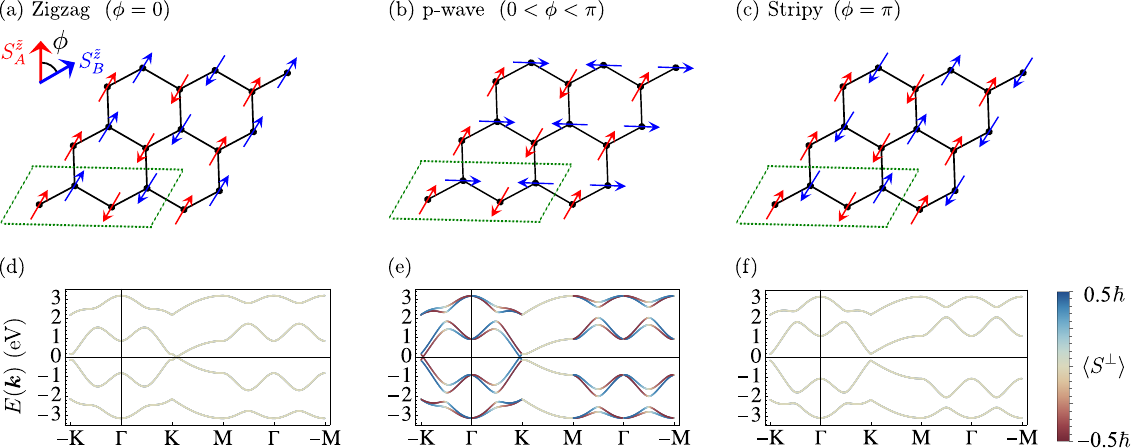}
	\caption{Spin configurations of (a) the collinear zigzag state, (b) the $p$-wave state at canting angle $\phi = 1.22$, and (c) the collinear stripy state. For all three states, the magnetic unit cell, indicated by the dashed green outline, is doubled relative to the crystallographic unit cell. (d)-(f) Band spectra of the model in Eq.~(\ref{s_eq:kondo}) with parameters $t = 1$ eV and $J_d = 0.6$ eV. The bands remain degenerate in the collinear zigzag and stripy states, whereas the $p$-wave state exhibits a characteristic spin splitting.}
	\label{s_fig_5}
\end{figure}

To illustrate the distinct band splitting in these three states, we introduce a minimal tight-binding model. The model is given by
\bea
H_{\text{p}} = t \sum_{\langle i j \rangle} c_{i}^\dagger c_{j}
+ J_d \sum_i c_{i}^\dagger [\hat{\bs{m}}_i \!\cdot\! \bs{\sigma}] c_{i},
\label{s_eq:kondo}
\eea
which is identical to that used in the main text, except that the spin texture is taken from the three configurations shown in Fig.~\ref{s_fig_5}(a)-(c). Here, $\hat{\bs{m}}_i$ denotes the site-dependent local spin polarization, and $J_d$ quantifies the coupling between the itinerant electrons and the localized moments. The electronic band structure of Eq.~(\ref{s_eq:kondo}) is shown in Fig.~\ref{s_fig_5}(d)-(f), plotted within the first Brillouin zone of the crystallographic unit cell; the number of bands is doubled due to zone folding. The resulting band dispersion exhibits a clear $p$-wave spin splitting in the $p$-wave state, whereas such splitting is absent in the collinear zigzag and stripy states.

\subsection{S4. Classical spin model for case (ii)}
\begin{figure}[h!]
	\includegraphics[width = 0.68\columnwidth]{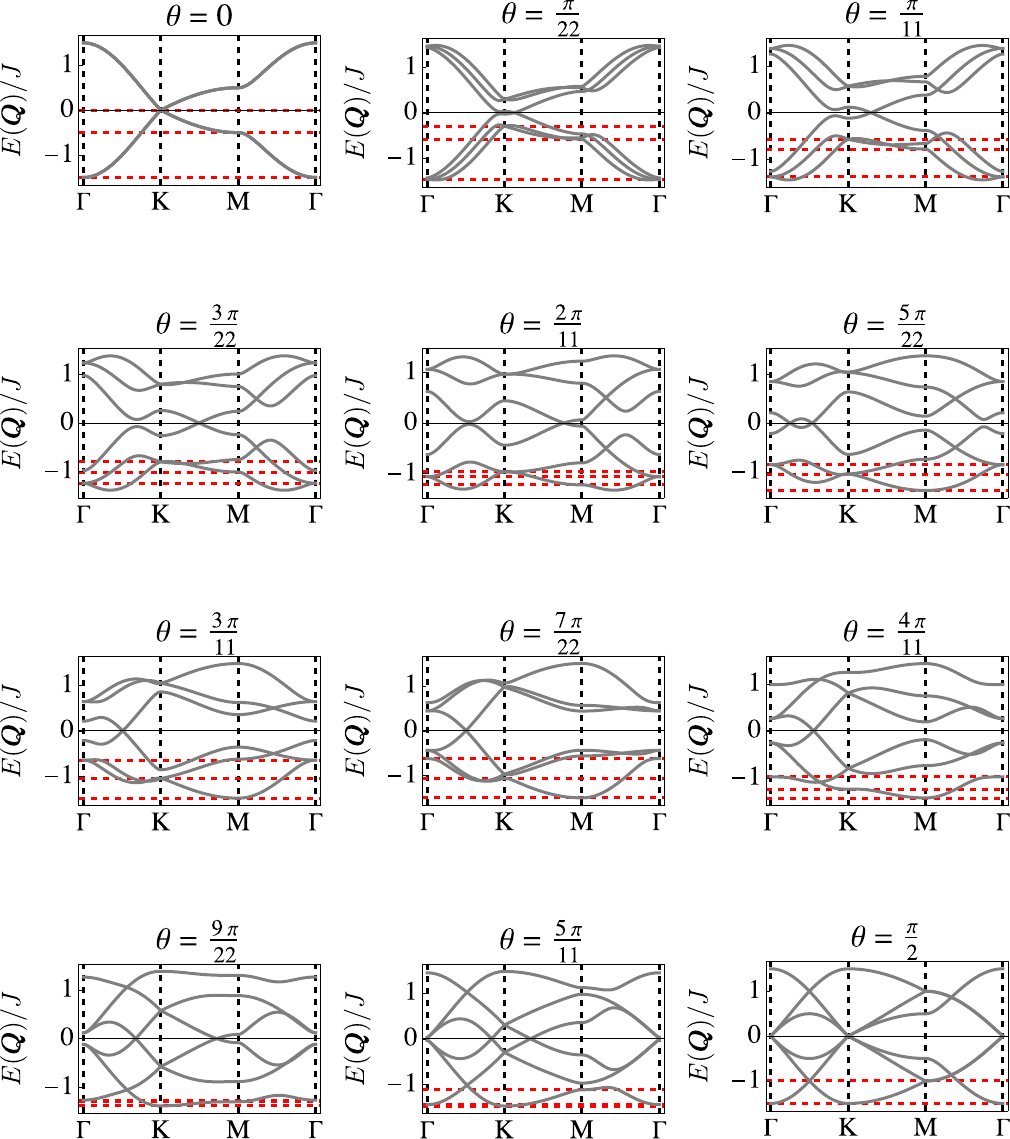}
	\caption{Eigenspectrum of the LT matrix for case (ii). For reference, the lowest eigenvalue at the high-symmetry points $\Gamma$, K, and M is indicated by three red dashed horizontal lines in each panel.}
	\label{s_fig_6}
\end{figure}
Here, we focus on the case (ii), where the $\hat{\bs d}_{\gamma}$ is written as
\bea
\hat{\bs d}_{x} = (1,0,0), \hat{\bs d}_{y} = C_{3z}\hat{\bs d}_{x}, \hat{\bs d}_{z} = C^2_{3z}\hat{\bs d}_{x}.
\label{s_eq:d_2}
\eea
The corresponding LT interaction matrix is given by :
\bea
\nonumber
H_\text{LT}(\bs Q) = \frac{J}{2}\left(
\begin{array}{cc}
	0 & h(\bs Q) \\
	h^T(-\bs Q) & 0
\end{array}
\right),
\eea
\bea
\nonumber
\footnotesize{
h(\bs Q)=
\left(
\begin{array}{ccc}
	\frac14\!\Big[
	4e^{i\bs n_x\bs Q}
	+e^{i\bs n_y\bs Q}(1+3J_\text{H})
	+e^{i\bs n_z\bs Q}(1+3J_\text{H})
	\Big]
	&
	\frac{\sqrt3}{4}\,\big(e^{i\bs n_y\bs Q}-e^{i\bs n_z\bs Q}\big)\,(-1+J_\text{H})
	&
	\frac{\sqrt3}{2}\,J_\text{D}\,\big(e^{i\bs n_y\bs Q}-e^{i\bs n_z\bs Q}\big)
	\\[10pt]
	\frac{\sqrt3}{4}\,\big(e^{i\bs n_y\bs Q}-e^{i\bs n_z\bs Q}\big)\,(-1+J_\text{H})
	&
	e^{i\bs n_x\bs Q}\,J_\text{H}
	+\frac14\big(e^{i\bs n_y\bs Q}+e^{i\bs n_z\bs Q}\big)(3+J_\text{H})
	&
	\frac12\,J_\text{D}\,\big(-2e^{i\bs n_x\bs Q}+e^{i\bs n_y\bs Q}+e^{i\bs n_z\bs Q}\big)
	\\[10pt]
	-\frac{\sqrt3}{2}\,J_\text{D}\,\big(e^{i\bs n_y\bs Q}-e^{i\bs n_z\bs Q}\big)
	&
	-\frac12\,J_\text{D}\,\big(-2e^{i\bs n_x\bs Q}+e^{i\bs n_y\bs Q}+e^{i\bs n_z\bs Q}\big)
	&
	\big(e^{i\bs n_x\bs Q}+e^{i\bs n_y\bs Q}+e^{i\bs n_z\bs Q}\big)\,J_\text{H}
\end{array}
\right)
}
\\
\eea
where $ J_\text{H}=\cos 2\theta$ and $J_\text{D}=\sin 2\theta$. The eigenspectrum is shown in Fig.~\ref{s_fig_6} for several representative values of $\theta$. Similar to case (i), for $0.67 \le \theta < 1.25$ the minimum of the lowest LT band lies at the $M$ points, and each corresponding eigenmode individually satisfies the hard-spin constraint. In this range, the single-$\bs Q$ $p$-wave magnet and the multi-$\bs Q$ noncoplanar states are energetically degenerate at the classical level.

For $1.25 \le \theta < \pi/2$, the minimum of the lowest LT band shifts to the $K$ points, and the system stabilizes a coplanar $120^\circ$ state. Fig.~\ref{s_fig_7}(a) displays the common-origin spin configuration obtained from iterative minimization. Spins on each sublattice form a $120^\circ$ pattern in the global $xy$-plane. The relative angle between the two sublattices is not fixed at the classical level. To characterize the $120^\circ$ state, we compute the triangular-lattice nearest-neighbor correlators separately on each sublattice (A/B) in a site-dependent local frame: $\langle S^{\tilde z}_i S^{\tilde z}_j\rangle$ and $\langle S^{\tilde z}_i S^{\tilde x}_j\rangle$ [Figs.~\ref{s_fig_7}(b–e)]. For every site $i$, the local axis $S^{\tilde z}_i$ is defined to be parallel to the spin-polarization direction at that site. In this convention, the mixed component vanishes, $\langle S^{\tilde z}_i S^{\tilde x}_j\rangle = 0$, on both sublattices [Figs.~\ref{s_fig_7}(c,e)], establishing that the state is coplanar in the global frame. Moreover, the correlator $\langle S^{\tilde z}_i S^{\tilde z}_j\rangle = 1$ on triangular nearest neighbors within each sublattice [Figs.~\ref{s_fig_7}(b,d)] demonstrates rigid alignment along the site-wise local polarization.

\begin{figure}[]
	\includegraphics[width = 0.65\columnwidth]{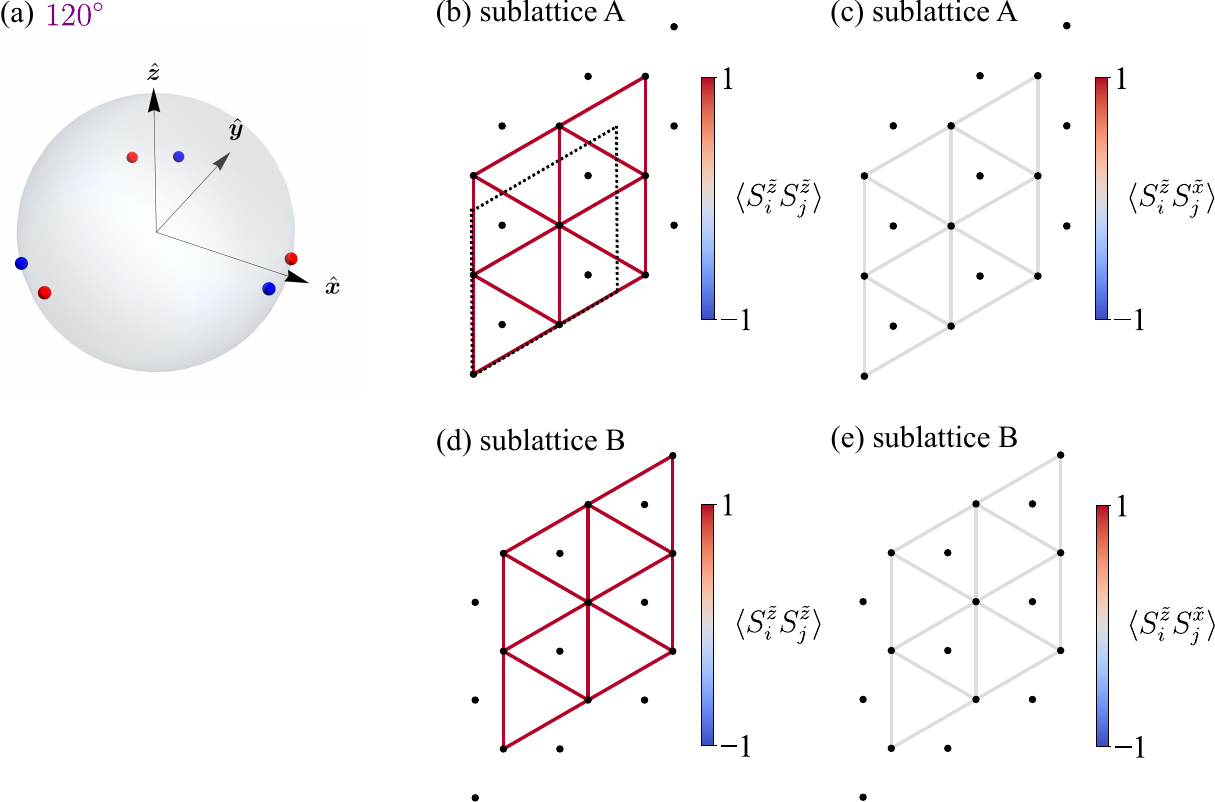}
	\caption{(a) Common-origin spin configuration of the $120^\circ$ state. Blue/red denote sublattices A/B. (b,c) Nearest-neighbor correlators on sublattice A: $\langle S^{\tilde z}_i S^{\tilde z}_j\rangle$ and $\langle S^{\tilde z}_i S^{\tilde x}_j\rangle$. The local axis $S^{\tilde z}_i$ is defined to be parallel to the spin-polarization direction of site $i$. (d,e) Nearest-neighbor correlators on sublattice B.}
	\label{s_fig_7}
\end{figure}

\subsection{S5. Quantum spin model for case (ii)}
For case (ii), we perform the same convergence analysis by monitoring the ground-state energy density as a function of the bond dimension $\chi$. To capture the relevant commensurate orders, we use an infinite-cylinder geometry with a unit cell of size $L_x\times L_y\times2=6\times3\times2$. This accommodates both the $p$-wave magnet and the coplanar $120^\circ$ state; possible incommensurate spiral states are disfavored by the finite unit-cell size.

At $\theta=\pi/4$, the ground state is the $p$-wave magnet. To examine the convergence of the iDMRG calculations, we analyze the scaling of the ground-state energy per site as a function of the bond dimension $\chi$. Fig.~\ref{s_fig_8}(a) shows the corresponding extrapolation in $1/\chi$, which exhibits a nearly linear trend and indicates good convergence. We perform the same analysis at $\theta=5\pi/12$, where the system stabilizes the $120^\circ$ state. As shown in Fig.~\ref{s_fig_8}(b), the ground-state energy per site again varies linearly with $1/\chi$, demonstrating that the iDMRG results are numerically well controlled for case (ii) as well.

For the case $\theta = 5\pi/12$, Figs.~\ref{s_fig_8}(c-f) further show the triangular-lattice nearest-neighbor correlators, computed separately on sublattices A and B in the site-dependent local frame. We evaluate $\langle S^{\tilde z}_i S^{\tilde z}_j\rangle$ and $\langle S^{\tilde z}_i S^{\tilde x}_j\rangle$ following the same procedure used in the classical analysis. One clearly sees that the mixed component $\langle S^{\tilde z}_i S^{\tilde x}_j\rangle$ vanishes, while $\langle S^{\tilde z}_i S^{\tilde z}_j\rangle$ remains finite and spatially uniform. This behavior unambiguously confirms that the state is the coplanar $120^\circ$ state.

\begin{figure}[]
	\includegraphics[width = 0.6\columnwidth]{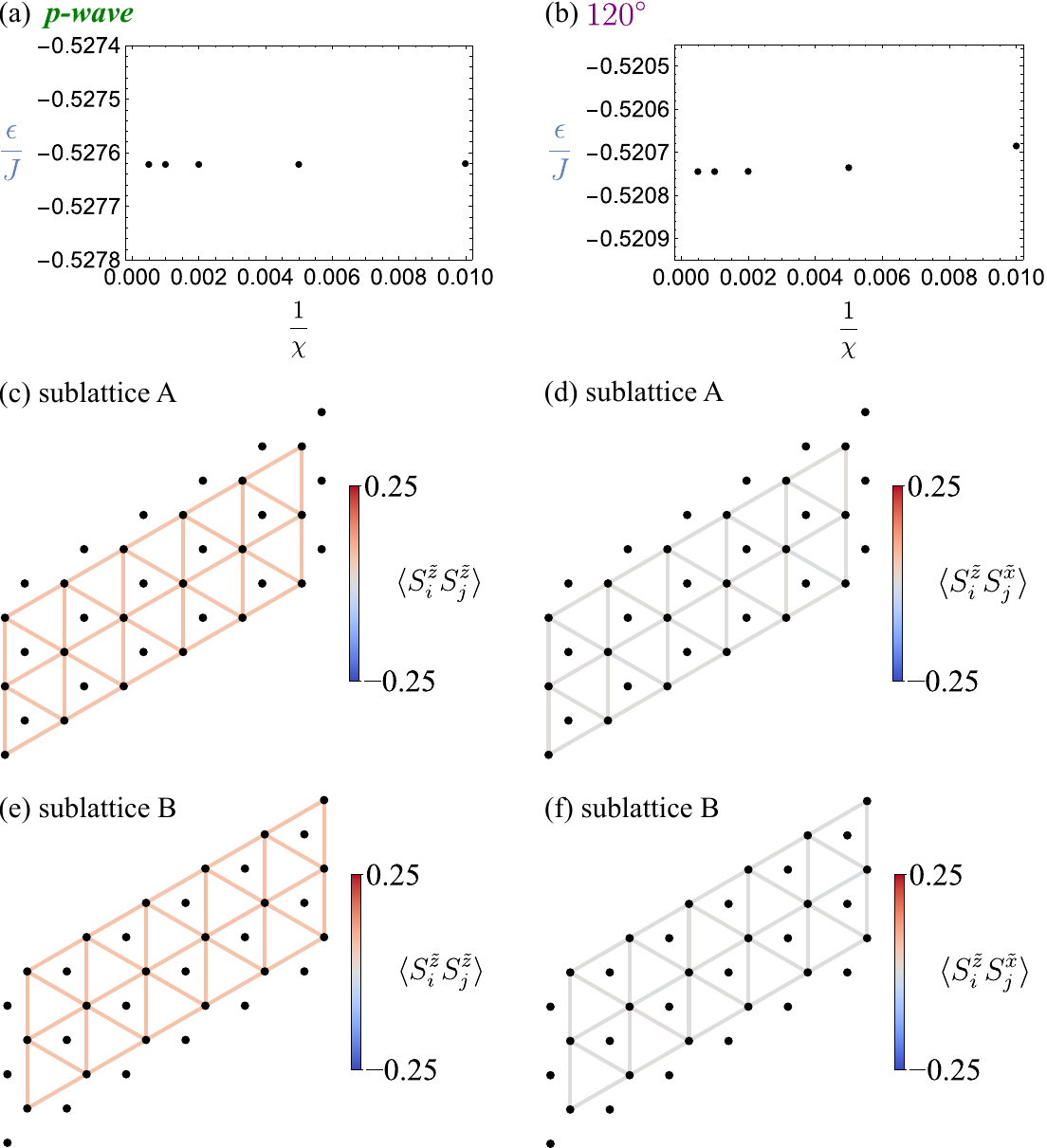}
	\caption{(a,b) Ground–state energy per site as a function of $1/\chi$ for $\theta=\pi/4$ and $\theta=5\pi/12$, respectively. The nearly linear dependence indicates that the iDMRG calculations are well converged. (c–f) Nearest–neighbor correlators for the $120^\circ$ state at $\theta=5\pi/12$, evaluated in the site–dependent local frame. The mixed component $\langle S^{\tilde z}_i S^{\tilde x}_j\rangle$ vanishes while $\langle S^{\tilde z}_i S^{\tilde z}_j\rangle$ is finite and uniform, confirming the coplanar $120^\circ$ character.}
	\label{s_fig_8}
\end{figure}

\end{widetext}

\bibliography{p_spin_bib.bib}

\end{document}